\documentclass[twoside]{dis08}
\usepackage[latin1]{inputenc}
\usepackage[dvips]{graphicx,epsfig,color}
\usepackage{wrapfig,rotating}
\usepackage{amssymb,amsmath,array}
\def\GeV{{\rm GeV}}

\begin{document}
\title{The Longitudinal Structure Function at HERA}

\author{R.S. Thorne\footnote{Royal Society University Research Fellow} 
%
%
\vspace{.3cm}\\
%
Department of Physics and Astronomy, University College London,
WC1E 6BT, UK}

\maketitle

\begin{abstract}
I investigate the theoretical uncertainties on the predictions for the 
longitudinal structure function $F_L(x,Q^2)$. I compare the predictions 
using fixed-order perturbative QCD, higher twist corrections, 
small-$x$ resummations and the dipole picture. I compare the various 
predictions to  the recent HERA measurements and examine how the data still 
to be analysed may discriminate between the approaches. 
\end{abstract}

\section{Introduction}

I consider the impact of a measurement of $F_L(x,Q^2)$ \cite{slides}.
Until recently we have been limited to consistency checks on the 
relationship between 
$F_2(x,Q^2)$ and $F_L(x,Q^2)$ at high $y$, where both contribute
to the total cross-section $\tilde \sigma(x,Q^2)=F_2(x,Q^2)
-y^2/(1+(1-y)^2)F_L(x,Q^2)$. Extracting $F_L$ {\em data} requires an 
extrapolation in $y$ making some theory assumptions. 
It is more useful to fit directly
to  $\tilde \sigma(x,Q^2)$ where 
there is a turn-over at the highest $y$. Previous studies have 
sometimes  found that at NLO the turn-over is too small, but better at NNLO
due to large corrections to $F_L(x,Q^2)$ \cite{MSTFL}. 
However, the precision of such 
studies is limited, and can be affected by systematics, e.g. the 
photo-production background uncertainty. Using the final low-energy run 
at HERA, a direct measurement of $F_L(x,Q^2)$ at HERA is  
possible \cite{H1, ZEUS}. 

\begin{wrapfigure}{r}{0.42\columnwidth}
\vspace{-0.5cm}
\centerline{\includegraphics[width=0.38\columnwidth]{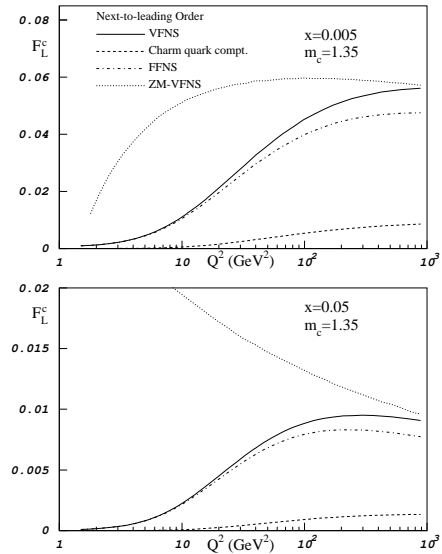}}
\vspace{-0.3cm}
\caption{An illustration of the charm contribution to $F_L(x,Q^2)$ in a 
general-mass variable flavour number scheme.}\label{Fig:flvfns}
\end{wrapfigure}

Here I outline  the implications 
and the predictions. The measurement of $F_L(x,Q^2)$ gives an independent 
test of the gluon distribution at low $x$ to accompany that determined
from $d F_2(x,Q^2)/ d \ln Q^2$.  However, it is also  a direct test of 
success of alternative theories in QCD.  These are slightly different issues.
It is not obvious for which $F_L(x,Q^2)$ has more discriminating 
power. First I will discuss the predictions at LO, 
NLO and NNLO perturbation theory. 
I briefly highlight the issue of heavy flavours, since this is more 
important for $F_L(x,Q^2)$ than for $F_2(x,Q^2)$ at low orders. The total
structure function is dominated by $C_{Lg}(\alpha_S,x)\otimes g(x,Q^2)$
contributions. In the massless quark approximation charge weighting means 
$F^c_L(x,Q^2)$is nearly $40\%$ of the total. 
However, there is a large massive quark suppression in heavy flavour 
coefficient functions. $F^c_L(x,Q^2)$ is suppressed by a factor of $v^3$ where 
$v= 1-\frac{4m_c^2 z}{Q^2 (1-z)}$ is the velocity of the heavy quark in the 
centre-of-mass frame. $Q^2 >> m_c^2$ before the massless limit starts 
to apply, as shown in Fig. \ref{Fig:flvfns}. 

\begin{figure}
\vspace{0.1cm}
\centerline{\includegraphics[width=0.45\columnwidth]{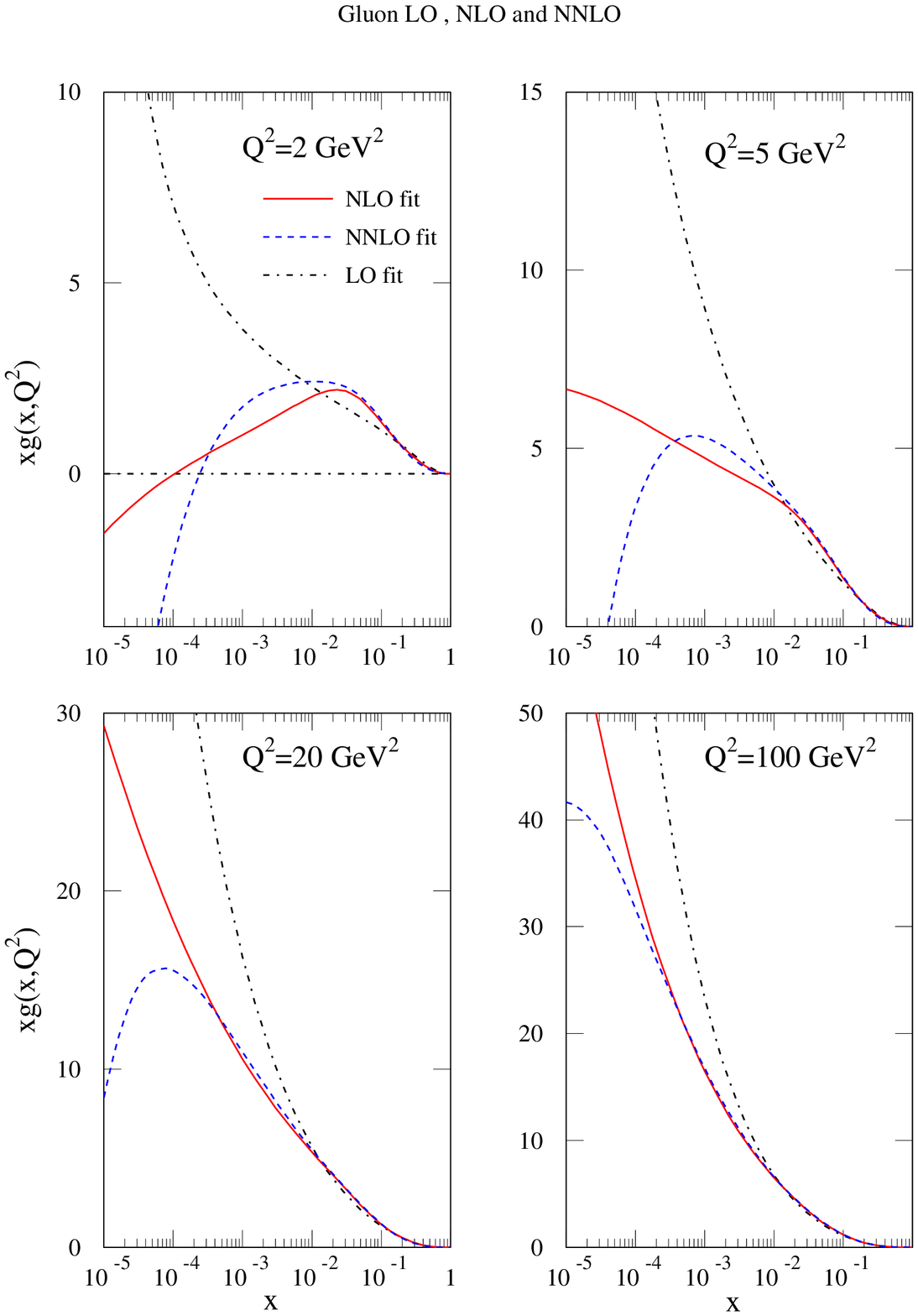}
\hspace{0.8cm}\includegraphics[width=0.45\columnwidth]{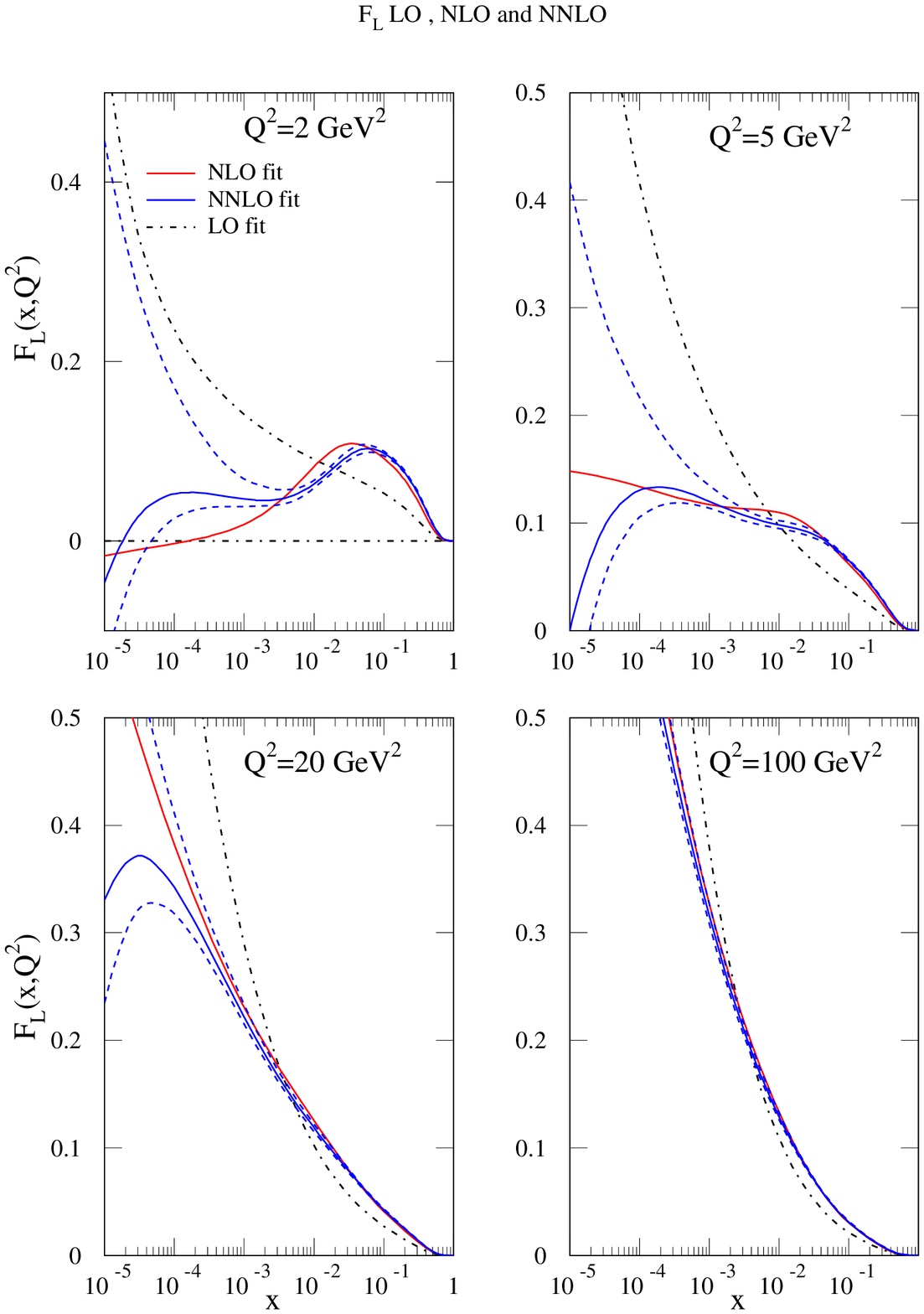}}
\vspace{-0.2cm}
\caption{The gluon distribution (left) and longitudinal structure function
(right) at LO, NLO and NNLO using preliminary MSTW08 
pdfs \cite{MSTW08}.}\label{Fig:partons}
\vspace{-0.4cm}
\end{figure}

At small $x$ the large order-by-order change in the splitting functions,
particularly $P_{qg}$, 
leads to a large variation in the gluon extracted from a global fit, shown in 
the left of Fig. \ref{Fig:partons}. 
However, the NNLO 
${\cal O}(\alpha_s^3)$ longitudinal coefficient function $C^3_{Lg}(x)$ 
\cite{NNLOFL} has a 
large positive contribution at small $x$, and this counters the decrease in 
the small-$x$ gluon. The predictions for $F_L(x,Q^2)$ at LO, NLO and NNLO 
are shown in the right of Fig. \ref{Fig:partons}. 
The $F_L(x,Q^2)$ prediction is more stable than the  gluon at 
small $x$. The uncertainty (shown only for NNLO) becomes enormous as $x$ 
decreases below $0.0001$, but at the lowest $Q^2$ the NLO and NNLO predictions 
are discrepant in some regions. The LO prediction is far larger than either, 
reflecting the huge correction in  the small-$x$ gluon going to NLO. 

\section{Beyond fixed order}

There are various potentially large corrections beyond fixed-order 
perturbation theory.
It is possible there is a large higher twist contribution from renormalons 
in the quark sector. For 
$F_2(x,Q^2)$ the renormalon calculation of higher twist
dies away at small $x$ (due to satisfying the Adler sum rule).
It is a completely different picture for $F_L(x,Q^2)$. At small $x$
the contribution is proportional to the quark distributions, i.e. 
$F_L^{HT}(x,Q^2) \propto F_2(x,Q^2)$.
The explicit renormalon calculation \cite{Stein} gives
\begin{equation}
F^{HT}_L(x,Q^2) = \frac{A}{Q^2}(\delta(1-x)-2x^3)\otimes \sum_f Q_f^2 
q_f(x,Q^2),\nonumber
\end{equation}
where $A \approx \frac{4C_f\exp(5/3)}{\beta_0} \Lambda_{QCD}^2
\approx 0.4-0.8\,\GeV^2.$
It should be stressed that this effect is nothing to do with the 
gluon distribution, and is
not part of the higher twist contribution included in the dipole approach. 
At small $x$ the correction becomes effectively
\begin{equation}
F^{HT}_L(x,Q^2) = \frac{A}{3Q^2}\delta(1-x)\otimes \sum_f Q_f^2 
q_f(x,Q^2) \approx \frac{A}{3Q^2}F_2(x,Q^2),
\nonumber
\end{equation}
and is $\sim 0.1$ for $x=0.0001$ and $Q^2=2\,\GeV^2$. 

\begin{figure}[h]
\vspace{-1.7cm}
\centerline{\includegraphics[width=0.4\columnwidth]{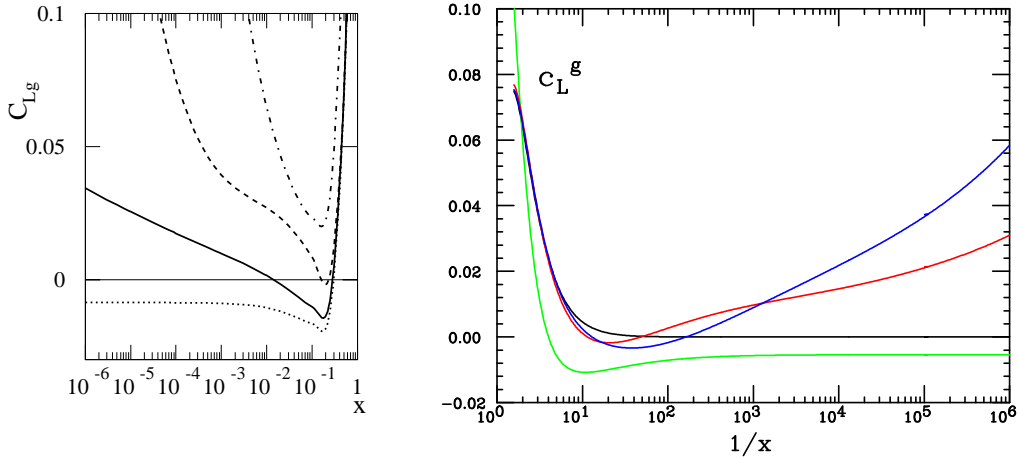}
\hspace{0.3cm}\includegraphics[width=0.55\columnwidth]{thorne_robert_2.fig5.ps}}
\vspace{-0.3cm}
\caption{The longitudinal coefficient function $C_{L,g}$
calculated using the approaches in \cite{WTres} (left, solid line), 
and \cite{ABFres} (right, blue line).}
\label{Fig:resumclg}
\end{figure}

If the small-$x$ NNLO correction is itself rather large, might not 
higher orders still be important? There are  
leading $\ln(1/x)$ terms of the form $P_{gg}(x) 
\sim \frac{\alpha_S^n \ln^{n-1}(1/x)}{x}, 
P_{qg}(x) \sim \frac{\alpha_S^n \ln^{n-2}(1/x)}{x}$ and 
$C_{Lg}(x) \sim \frac{\alpha_S^n \ln^{n-2}(1/x)}{x}$.
A fit which performs a double resummation 
of leading $\ln(1/x)$ and $\beta_0$ terms leads to a better fit
to small-$x$ data than a conventional perturbative fit \cite{WTres}.
The gluon distribution from this resummed fit is larger at small $x$ and 
$Q^2$ than NLO or NNLO,  
and this is reflected also in the prediction for $F_L(x,Q^2)$.  
Similar approaches \cite{ABFres,CCSSres} all lead to rather comparable 
results for the calculated splitting functions, but only in \cite{WTres} 
has detailed phenomenological studies taken place. A comparison of 
the longitudinal coefficient functions from two approaches 
is shown in Fig. \ref{Fig:resumclg}. The two
results are clearly of the same form, so it is expected that 
a prediction for $F_L(x,Q^2)$ using the approach in 
\cite{ABFres} should be similar  to that produced in \cite{WTres}.  

Finally I consider the dipole picture \cite{dipole}. 
As with small-$x$ resummations this 
can be cast in the language of $f(x,k^2)$ -- the unintegrated gluon 
distribution -- which is directly related to the dipole-proton cross-section. 
The structure functions are obtained by convoluting this dipole cross-section
with the wave-functions for the photon to fluctuate into a quark-antiquark
pair.  This picture includes some of the resummation effects, 
and also higher twist contributions, and is designed to approach $Q^2=0$ 
smoothly. However, it misses quark and higher-$x$ contributions. Overall 
$F_L(x,Q^2)$ predicted in this approach is steeper at 
small $x$ than fixed order, and automatically stable at lowest $Q^2$,
see e.g. \cite{GBW}. The general features are rather
insensitive to whether saturation effects are included in the dipole 
cross-section.

\section{Predictions}

\begin{wrapfigure}{r}{0.65\columnwidth}
\vspace{-1.1cm}
\centerline{\includegraphics[width=0.65\columnwidth]{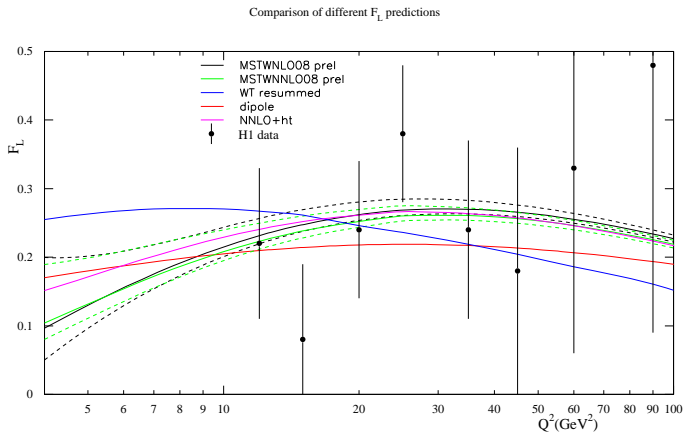}}
\vspace{0.2cm}
\caption{A comparison of the various predictions for $F_L(x,Q^2)$ 
with the HERA data.}\label{Fig:preds}
\end{wrapfigure}

I present the various predictions for $F_L(x,Q^2)$ along a line of 
$x=Q^2/35420$, which corresponds well to the current HERA measurements 
\cite{H1FL}, and to those yet to appear. The data are in good agreement 
with all the predictions. Along this line the NLO and NNLO 
predictions \cite{MSTW08} are very similar, and the higher 
twist corrections are slightly smaller than the pdf uncertainties at 
NLO and NNLO. The resummation prediction \cite{WTres} and dipole model 
prediction \cite{RTdip} (very similar in \cite{Wattdip}) have a different 
shape, and it is perhaps possible to distinguish the former at lower $Q^2$.  
At $Q^2 \geq 10\,\GeV^2$  the uncertainty on fixed order 
predictions is a few percent. An $F_L(x,Q^2)$ measurement will not add 
much to the direct constraint on the gluon. However, there 
may be deviations from
NLO/NNLO predictions of $20-30\%$ due to e.g. resummations or 
dipole models. 
For $Q^2 \leq 10\,\GeV^2$ the uncertainty in NLO/NNLO 
predictions for $F_L(x,Q^2)$ due to the gluon uncertainty increases to
$> 20\%$. A good measurement of $F_L(x,Q^2)$ here will automatically
improve the gluon determination. 
Resummations/dipole models suggest a higher low-$Q^2$
$F_L(x,Q^2)$ by an absolute value of up to $0.15$ -- well
outside the fixed-order uncertainties. A good measurement of 
$F_L(x,Q^2)$ will start to discriminate between theories.


\begin{footnotesize}

\end{footnotesize}

\begin{thebibliography}{99}

\bibitem{slides} Slides: \\ 
\verb$http://indico.cern.ch/contributionDisplay.py?contribId=93&sessionId=17&confId=24657$

\bibitem{MSTFL}
  A.~D.~Martin, W.~J.~Stirling and R.~S.~Thorne,
  Phys.\ Lett.\  B {\bf 635} (2006) 305
  [arXiv:hep-ph/0601247].



\bibitem{H1}
 B. Antunovic, these proceedings; V. Chekelian, these proceedings.

\bibitem{ZEUS} D. Kollar, these proceedings. 

\bibitem{NNLOFL} S.~Moch, J.~A.~M.~Vermaseren and A.~Vogt,
  Phys.\ Lett.\  B {\bf 606} (2005) 123
  [arXiv:hep-ph/0411112]; J.~A.~M.~Vermaseren, A.~Vogt and S.~Moch,
  Nucl.\ Phys.\  B {\bf 724} (2005) 3
  [arXiv:hep-ph/0504242].

\bibitem{MSTW08}
 G.~Watt, A.~D.~Martin, W.~J.~Stirling and R.~S.~Thorne,
  arXiv:0806.4890 [hep-ph].

\bibitem{Stein}
E.~Stein, {\it et al.},
  Phys.\ Lett.\  B {\bf 376} (1996) 177
  [arXiv:hep-ph/9601356].

\bibitem{WTres}
C.~D.~White and R.~S.~Thorne,
  Phys.\ Rev.\  D {\bf 75} (2007) 034005
  [arXiv:hep-ph/0611204].

\bibitem{ABFres}
  G.~Altarelli, R.~D.~Ball and S.~Forte,
  Nucl.\ Phys.\  B {\bf 799} (2008) 199
  [arXiv:0802.0032 [hep-ph]].

\bibitem{CCSSres}
M.~Ciafaloni, {\it et al.},
  JHEP {\bf 0708} (2007) 046
  [arXiv:0707.1453 [hep-ph]].

\bibitem{dipole}
L.~L. Frankfurt and M.~I. Strikman,
Phys.\ Rept.\  {\bf 160}, (1988) 235; A.~H. Mueller,
Nucl.\ Phys.\ {\bf B335}, (1990) 115; N.~N. Nikolaev and B.~G. Zakharov,
Z.\ Phys.\ {\bf C49}, (1991) 607.

\bibitem{GBW}
K. Golec-Biernat and M. Wusthoff,
%
Phys.\ Rev.\ {\bf D59}, (1999) 014017.

\bibitem{H1FL}
  F.~D.~Aaron {\it et al.}  [H1 Collaboration],
  Phys.\ Lett.\  B {\bf 665} (2008) 139
  [arXiv:0805.2809 [hep-ex]].

\bibitem{RTdip}
 R.~S.~Thorne,
  Phys.\ Rev.\  D {\bf 71} (2005) 054024
  [arXiv:hep-ph/0501124].

\bibitem{Wattdip}
 G.~Watt and H.~Kowalski,
  Phys.\ Rev.\  D {\bf 78} (2008) 014016
  [arXiv:0712.2670 [hep-ph]].





\end{thebibliography}
\end{document}